\documentclass[prd,aps,showkeys,superscriptaddress,two column]{revtex4-1}

\usepackage{amsmath}
\usepackage{amssymb}
\usepackage{amsthm}
\usepackage{mathrsfs}
\usepackage{graphicx}
\usepackage{fancyhdr}
\usepackage{array}
\usepackage{hyperref}
\usepackage{simplewick}
\usepackage{latexsym}
\usepackage[all]{xy}
\usepackage{eufrak}
\usepackage{euscript}
\usepackage{enumerate}
\usepackage{dsfont}
\usepackage{slashed}

\newcommand{\be}{\begin{equation}}
\newcommand{\ee}{\end{equation}}
\newcommand{\bs}{\begin{split}}
\newcommand{\es}{\end{split}}

\def \v {\vec}

\begin{document}



\title{$U(1)$ effective confinement theory from $SU(2)$ restricted gauge theory via the Julia-Toulouse Approach}

\author{L. S. Grigorio}
\email{leogrigorio@if.ufrj.br}
\affiliation{Instituto de F\'\i sica, Universidade Federal do Rio de Janeiro,
\\21941-972, Rio de Janeiro, Brazil}
\affiliation{Centro Federal de Educa\c{c}\~ao Tecnol\'ogica Celso Suckow da Fonseca,
\\28635-000, Nova Friburgo, Brazil}

\author{M. S. Guimaraes}
\email{msguimaraes@uerj.br}
\affiliation{Departamento de F\'\i sica Te\'orica, Instituto de F\'\i sica, UERJ - Universidade do
Estado do Rio de Janeiro, Rua S\~ ao Francisco Xavier 524, 20550-013 Maracan\~ a, Rio de Janeiro,
Brazil}

\author{W. Oliveira}
\email{wilson@fisica.ufjf.br}
\affiliation{Departamento de F\'\i sica, ICE,
\\Universidade Federal de Juiz de Fora,
\\36036-330, Juiz de Fora, MG, Brazil}

\author{R. Rougemont}
\email{romulo@if.ufrj.br}
\affiliation{Instituto de F\'\i sica, Universidade Federal do Rio de Janeiro,
\\21941-972, Rio de Janeiro, Brazil}

\author{C. Wotzasek}
\email{clovis@if.ufrj.br}
\affiliation{Instituto de F\'\i sica, Universidade Federal do Rio de Janeiro,
\\21941-972, Rio de Janeiro, Brazil}

\date{\today}

\begin{abstract}
We derive an $U(1)$ effective theory of color confinement by applying the so-called Julia-Toulouse Approach for defects
condensation to the $SU(2)$ restricted gauge theory defined by means of the Cho decomposition of the non-abelian
connection. Cho's geometric construction naturally displays the topological degrees of freedom of the theory and can be
used to put the Yang-Mills action into an abelianized form under certain conditions. On the other hand, the use of the
Julia-Toulouse prescription to deal with the monopole condensation leads to an effective action describing the phase whose
dynamics is dominated by the magnetic condensate. The effective theory we found describes the interaction between external
electric currents displaying a short-range Yukawa interaction plus a linear confinement term that governs the long
distance physics.
\end{abstract}


\keywords{Topological defects, monopoles, confinement.}


\maketitle


\section{Introduction}
\label{sec:introduction}

It is known that due to the Meissner effect the usual superconductors should confine magnetic monopoles. This point led to
the conjecture that the QCD vacuum could be a condensate of chromomagnetic monopoles, a dual superconductor as originally
proposed in \cite{intro}. Such a chromomagnetic condensate should be responsible for the dual Meissner effect that is
expected to lead to the confinement of color charges immersed in this medium. In dual superconductor models of color
confinement, magnetic monopoles usually appear as topological defects in points of the space where the abelian projection
becomes singular. For a review, see for example \cite{ripka}.

In this Letter we follow a different path to reveal the magnetic monopole condensate in the pure $SU(2)$ gauge theory.
First, instead of just writing down an effective Dual Abelian Higgs Model compatible with the residual gauge symmetry
obtained from the abelian projection of the $SU(2)$ gauge theory, we use the so-called Cho decomposition \cite{cho} of
the $SU(2)$ connection, which has the feature of explicitly exposing the abelian component of the non-abelian connection
revealing the topological structures of the theory without resorting to any singular gauge fixing procedure like the
abelian projection. We work with the subsector of the complete gauge theory called restricted gauge theory which contains
the full $SU(2)$ gauge degrees and is expected to be responsible for the color confinement, as claimed in \cite{cho}.

Next we fix the so-called magnetic gauge where we show that the action acquires the form of the Maxwell theory
minimally coupled to external chromoelectric charges and non-minimally coupled to chromomagnetic monopoles. At this
point we are in position to apply the Julia-Toulouse Approach (JTA) for defects condensation \cite{JT} as generalized
by some of us in \cite{csm, dafdc}. This is a prescription used to obtain an effective theory for a phase with condensed defects
starting from the theory defined in the phase where the defects are diluted - and exploit the consequences of the
monopole condensation. As the result, the effective theory describing the interaction between the chromoeletric charges
immersed in the chromomagnetic condensate features two parts:
\begin{itemize}
\item the first one is a Yukawa-like term that dominates the short-range physics in the magnetic
condensate - a typical feature of the confinement scenarios due to monopole condensation;
\item the second one describes, in the static case, a linear potential in the interquarks
separation, thus being responsible for the chromoeletric confinement that dominates the physics at
large distances.
\end{itemize}

\section{Setting the problem}
\label{sec:setting}

We begin with a brief review of the Cho decomposition of the $SU(2)$ connection. The starting point
is the introduction of a unitary color triplet, $\hat{n}:\mathbb{R}^{1,3}_{(\mbox{spacetime})}
\rightarrow S^2\subset\mathbb{R}^3_{(\mbox{color})},\,x\mapsto\hat{n}(x)|\hat{n}^2(x)=1$, and the
definition of the so-called restricted connection, $\hat{A}_\mu$, which leaves $\hat{n}$ invariant
under parallel transport on the principal bundle \cite{cho},
\begin{equation}
\hat{D}_\mu\hat{n}:=\partial_\mu \hat{n} + g\hat{A}_\mu\times\hat{n}\equiv 0 \Rightarrow
\hat{A}_\mu=A_\mu\hat{n}-\frac{1}{g}\hat{n}\times\partial_\mu\hat{n}.
\label{eq:1}
\end{equation}

As we are going to see in a moment:
\begin{itemize}
\item the unitary triplet $\hat{n}$ selects the abelian direction in the internal color space for
each spacetime point;
\item $A_\mu$ transforms like an $U(1)$ connection;
\item the restricted connection $\hat{A}_\mu$ is already an $SU(2)$ connection.
\end{itemize}

Due to the fact that the space of connections is an affine space, a general $SU(2)$ connection,
$\v{A}_\mu$, can be obtained from the restricted connection, $\hat{A}_\mu$, by adding a field
$\v{X}_\mu$ that is orthogonal to $\hat{n}$ \cite{cho}. Thus, the general form of the Cho
decomposition of the $SU(2)$ connection is given by
\begin{align}
&\v{A}_\mu=A_\mu\hat{n}-\frac{1}{g}\hat{n}\times\partial_\mu\hat{n}+\v{X}_\mu,\nonumber\\
&\hat{n}^2=1\,\,\mbox{and}\,\,\hat{n}\cdot\v{X}_\mu=0.
\label{eq:2}
\end{align}

From the infinitesimal $SU(2)$ gauge transformation defined by
\begin{align}
&\delta\v{A}_\mu=\frac{1}{g} D_\mu\v{\omega}:=\frac{1}{g}(\partial_\mu\v{\omega}+g\v{A}_\mu\times
\v{\omega}),\nonumber\\
&\delta\hat{n}=-\v{\omega}\times\hat{n},
\label{eq:3}
\end{align}
it follows that
\begin{align}
&\delta A_\mu=\frac{1}{g}\hat{n}\cdot\partial_\mu\v{\omega},\nonumber\\
&\delta\hat{A}_\mu=\frac{1}{g}\hat{D}_\mu\v{\omega},\nonumber\\
&\delta\v{X}_\mu=-\v{\omega}\times\v{X}_\mu.
\label{eq:4}
\end{align}

We see from (\ref{eq:4}) that $A_\mu$ transforms like an $U(1)$ connection, being the abelian
component of the $SU(2)$ connection explicitly revealed by the Cho decomposition without any gauge
fixing procedure (like the abelian projection). Thus, we say that the unitary triplet field
$\hat{n}$ selects the abelian direction in the color space for each spacetime point. Furthermore,
we also see from (\ref{eq:3}) and (\ref{eq:4}) that the restricted connection, $\hat{A}_\mu$,
transforms like the general $SU(2)$ connection, $\v{A}_\mu$, since the restricted covariant derivative
is expressed (in the adjoint representation), like the general covariant derivative, in terms of the
$SU(2)$ structure constants $\epsilon_{abc}$. Hence, as anticipated, the restricted connection is
already an $SU(2)$ connection carrying all the gauge degrees (but not all the dynamical degrees) of
the non-abelian gauge theory, being $\v{X}_\mu$ a source term called the valence potential which
carries the remaining dynamical degrees of the theory.

We shall concentrate our attention from now on into the restricted connection, since it already
gives us an $SU(2)$ theory, whose properties we are interested in analyze in this Letter. In fact,
as claimed in \cite{cho}, the restricted gauge theory governs the subdynamics of the complete gauge
theory that characterizes the vacuum of the theory and would be responsible for the color confinement,
that is what we are looking for.

The restricted curvature tensor is given by
\begin{align}
\hat{F}_{\mu\nu}&:=\partial_\mu\hat{A}_\nu-\partial_\nu\hat{A}_\mu+g\hat{A}_\mu\times\hat{A}_\nu
\nonumber\\
&=(F_{\mu\nu}+H_{\mu\nu})\hat{n},
\label{eq:5}
\end{align}
where
\begin{align}
&F_{\mu\nu}:=\partial_\mu A_\nu-\partial_\nu A_\mu,\nonumber\\
&H_{\mu\nu}:=-\frac{1}{g}\hat{n}\cdot(\partial_\mu\hat{n}\times\partial_\nu\hat{n}).
\label{eq:6}
\end{align}

Parametrizing $\hat{n}$ over $S^2$ by the polar angle, $\theta$, and azimuthal
angle, $\varphi$, $\hat{n}=(\sin(\theta)\cos(\varphi),\sin(\theta)\sin
(\varphi),\cos(\theta))$, we rewrite the last equation as
\begin{equation}
{H}_{\mu\nu}=-\frac{1}{g}\sin(\theta)(\partial_\mu\theta\partial_\nu\varphi-\partial_\mu\varphi
\partial_\nu\theta).
\label{eq:7}
\end{equation}

At this point we go to the so-called magnetic gauge defined by fixing the local color vector
field $\hat{n}$ in the $\hat{z}$-direction in the internal color space \cite{cho}. In this
gauge, the curvature tensor is written as $\hat{F}_{\mu\nu}=(F_{\mu\nu}+H_{\mu\nu})\hat{z}$.
Defining the so-called magnetic potential by the expression
\begin{equation}
\tilde{C}_\mu:=\frac{1}{g}(\cos(\theta)\partial_\mu\varphi+\partial_\mu\gamma),
\label{eq:8}
\end{equation}
we see that we can rewrite $H_{\mu\nu}$ in the abelianized form,
\begin{equation}
H_{\mu\nu}=\partial_\mu\tilde{C}_\nu-\partial_\nu\tilde{C}_\mu.
\label{eq:9}
\end{equation}

The angle $\gamma$ is the third of the Euler angles used to define a general $SO(3)$
transformation that rotates $\hat{n}$ into $\hat{z}$ in $\mathbb{R}^3_{(\mbox{color})}$.

It is easy to see now that the restricted connection transforms like $\hat{A}_\mu\mapsto(A_\mu+
\tilde{C}_\mu)\hat{z}$ under the gauge transformation that leads us to the magnetic gauge.

The magnetic potential, $\tilde{C}_\mu$, describes the potential of a monopole, being singular over
its associated Dirac string \cite{dirac}, as we can easily see following the example discussed in
\cite{oxman}: if we consider $\gamma=-\varphi$, we have from (\ref{eq:8}) that
\begin{equation}
\tilde{C}_\mu=\frac{1}{g}(\cos(\theta)-1)\partial_\mu\varphi.
\label{eq:10}
\end{equation}

Since we have for the gradient in spherical coordinates that $\partial_0:=\partial_t\equiv
\partial/\partial t$, $\partial_1:=\partial_r\equiv\partial/\partial r$, $\partial_2:=
\partial_{\theta}\equiv(1/r)\partial/\partial\theta$ and $\partial_3:=\partial_{\varphi}\equiv
(1/r\sin(\theta))\partial/\partial\varphi$, we see from (\ref{eq:10}) that
\begin{equation}
\tilde{C}_\mu=\frac{1}{g}\frac{(\cos(\theta)-1)}{r\sin(\theta)}\delta_{\mu\varphi},
\label{eq:11}
\end{equation}
which is the monopole potential singular over the Dirac string arbitrarily placed (by the choice
made for $\gamma$) in the negative $\hat{z}$-axis ($\theta=\pi$).

This singularity is a gauge artifact and must not show up in the final expressions for the
physical observables. Hence, in order to define a regular finite action, we must subtract the
unphysical singularity that arises in the expression for $H_{\mu\nu}$ due to the flux tube
inside the Dirac string, introducing a $\delta$-distritution, $\Lambda_{\mu\nu}^M$, that localizes
the world surface spanned by the magnetic Dirac string and exactly cancels out the singularity in
$H_{\mu\nu}$, as discussed for example, in chapter 8 of \cite{mvf} and chapter 2 of \cite{ripka}.
This reasoning leads us to write the Lagrangian density for the restricted theory as:
\begin{align}
\mathcal{L}&=-\frac{1}{4}\hat{F}^2_{\mu\nu} \nonumber\\
&=-\frac{1}{4}F_{\mu\nu}^2-\frac{1}{2}F_{\mu\nu}(H^{\mu\nu}-\Lambda^{\mu\nu}_M)-
\frac{1}{4}(H_{\mu\nu}-\Lambda_{\mu\nu}^M)^2.
\label{eq:12}
\end{align}

Now we follow the reasoning presented in the model reviewed in section 5.2 of \cite{ripka} and
minimally couple external chromoeletric currents, $j_\mu$ (described by electric Dirac strings,
$\tilde{\Lambda}_{\mu\nu}^E$, through the relation $j_\mu=\frac{1}{2}\epsilon_{\mu\nu\alpha\beta}
\partial^\nu\tilde{\Lambda}_E^{\alpha\beta}$), to the restricted potential expressed in the magnetic
gauge, $(A_\mu+\tilde{C}_\mu)$, obtaining the following Lagrangian density:
\begin{align}
\bar{\mathcal{L}}&=-\frac{1}{4}F_{\mu\nu}^2-\frac{1}{2}F_{\mu\nu}(H^{\mu\nu}-\Lambda^{\mu\nu}_M)-
\frac{1}{4}(H_{\mu\nu}-\Lambda_{\mu\nu}^M)^2+\nonumber\\
&-(A_\mu+\tilde{C}_\mu)j^\mu.
\label{eq:13}
\end{align}

We shall refer generically to the magnetic (electric) Dirac string and to its world surface as a
``magnetic (electric) Dirac brane".

\section{The Julia-Toulouse Approach for the monopole condensation and the $U(1)$ effective
theory of confinement}
\label{sec:JTA}

Notice that absorbing the singular monopole field $\tilde{C}_\mu$ into the regular abelian gluon field $A_\mu$
by redifining $(A_\mu+\tilde{C}_\mu)\mapsto A_\mu$, we can rewrite (\ref{eq:13}) in the following form:
\begin{align}
\bar{\mathcal{L}}=-\frac{1}{4}(F^{\mu\nu}-\Lambda^{\mu\nu}_M)^2-j_\mu A^\mu,
\label{eq:14}
\end{align}
where now the field $A_\mu$ is singular over the magnetic Dirac branes. Equation (\ref{eq:14}) describes the Maxwell
theory with the vector potential $A_\mu$ minimally coupled to electric currents and non-minimally coupled to monopoles.
We have exploited the monopole condensation phenomenon in this action and in its dual counterpart in great details in
\cite{dafdc}. However, notice that here the monopoles were not included in the theory by hand, instead they were naturally
revealed in the YM theory by the Cho decomposition of the non-abelian connection. Furthermore, the quantum theory associated
to (\ref{eq:14}) must be invariant under deformations of the unphysical Dirac strings. This is accomplished provided we
impose the non-abelian version of the Dirac quantization condition \cite{dirac, ripka}, $g\tilde{g}=4\pi n,\,n\in
\mathbb{Z}$, where $\frac{g}{2}$ (which is present in the electric string term, $\tilde{\Lambda}_{\mu\nu}^E:=\frac{g}{2}
\tilde{\delta}_{\mu\nu}(x;S_E)$, being $S_E$ the world surface of the electric Dirac string) is the $SU(2)$
chromoeletric charge of the quarks, being $g$ the QCD coupling constant, and $\tilde{g}=\frac{4\pi n}{g}$ (which is
present in the magnetic string term, $\Lambda_{\mu\nu}^M:=\tilde{g}\tilde{\delta}_{\mu\nu}(x;S_M)$, being $S_M$ the
world surface of the magnetic Dirac string) is the chromomagnetic charge of the monopoles in the $n$-th homotopy class
of the mapping $\Pi_2(SU(2)/U(1)\simeq S^2)=\mathbb{Z}$ defined by the unitary triplet $\hat{n}$ \cite{cho}.

We are now in position to apply the JTA to obtain an effective theory describing the phase where the monopoles are
condensed. In (\ref{eq:14}), the field $A_\mu$ is regular only over $\mathbb{R}^{1,3}_{(\mbox{spacetime})}\backslash
\mathcal{M}$, where $\mathcal{M}$ is the geometric place of the magnetic Dirac branes. As the monopoles proliferate,
the magnetic potential can only be defined over an increasingly smaller region in the space until we reach the critical
case where the monopoles proliferate occupying the whole space. In this case, $A_\mu$ can not be defined anywhere.
Equivalently, the Dirac branes of the condensing monopoles occupy the whole space and should be elevated to the field
category describing the long wavelength fluctuations of the condensate. The Julia-Toulouse procedure consists in the
observation that the regular physical combination $(F_{\mu\nu}-\Lambda_{\mu\nu}^M)$ should be taken as the fundamental
field $Y_{\mu\nu}$ describing the magnetic monopole condensate \cite{JT}. This becomes the magnetic equivalent to the
Stuckelberg procedure where the condensate field ``eats up" the gauge field to become massive. Notice that in doing so
we have effectively promoted the kinetic term for the 1-form gauge field describing the normal or diluted phase to a
mass term for the 2-form Kalb-Ramond field describing the monopole condensate in the condensed phase - this mass
generation accompanied by the rank-jump of the field describing the defects condensate is the main signature of the
JTA \cite{Gaete:2004dn, csm, dafdc}. Next, we must give dynamics to the 2-form describing the magnetic condensate
supplementing the action with a kinetic term for it which, usually, results from a Lorentz and gauge symmetry preserving
derivative expansion \cite{Gamboa:2008ne} and the outcome of such approach is the following effective theory for the
magnetic condensed phase \cite{JT, dafdc}:
\begin{align}
\bar{\mathcal{L}}_c&=\frac{1}{12}(\partial_\mu Y_{\alpha\beta}+\partial_\alpha Y_{\beta\mu}
+\partial_\beta Y_{\mu\alpha})^2+\frac{m_Y}{4}Y_{\mu\nu}\epsilon^{\mu\nu\alpha\beta}\tilde{\Lambda}^
E_{\alpha\beta}+\nonumber\\
&-\frac{m_Y^2}{4}Y_{\mu\nu}^2.
\label{eq:15}
\end{align}

In Minkowski spacetime the dual Kalb-Ramond field, $\tilde{Y}_{\mu\nu}:=\frac{1}{2!}\epsilon_{\mu\nu\alpha\beta}
Y^{\alpha\beta}$, implies the relations:
$$
\left\{\begin{array}{l}
\frac{1}{12}(\partial_\mu Y_{\alpha\beta}+\partial_\alpha Y_{\beta\mu}+\partial_\beta Y_{\mu\alpha})
^2=-\frac{1}{2}(\partial_\mu\tilde{Y}^{\mu\nu})^2\\
Y_{\mu\nu}^2=-\tilde{Y}_{\mu\nu}^2\\
\frac{1}{4}Y_{\mu\nu}\epsilon^{\mu\nu\alpha\beta}\tilde{\Lambda}^E_{\alpha\beta}=\frac{1}{2}
\tilde{Y}_{\mu\nu}\tilde{\Lambda}^{\mu\nu}_E\\
{Y}_{\mu\nu}=-\frac{1}{2}\epsilon_{\mu\nu\alpha\beta}\tilde{Y}^{\alpha\beta}
\end{array}\right.,
$$
such that in terms of $\tilde{Y}_{\mu\nu}$ the effective action describing the condensed phase is written as:
\begin{align}
\bar{S}_{eff}[\tilde{Y}_{\mu\nu}]&=\int d^4x\left[-\frac{1}{2}(\partial_\mu\tilde{Y}^{\mu\nu})^2
+\frac{m_Y^2}{4}\tilde{Y}_{\mu\nu}^2+\frac{m_Y}{2}\tilde{Y}_{\mu\nu}\tilde{\Lambda}_E^{\mu\nu}\right].
\label{eq:16}
\end{align}

Equation (\ref{eq:16}) is the result of the application of the JTA, as formulated in the relativistic field theory context
by Quevedo-Trugenberger, to the Maxwell theory. This equation, however, features an undesirable point: it is not invariant
under deformations of the unphysical electric Dirac strings. If we deform $S_E\mapsto S_E',\,\partial S_E=\partial
S_E'$, where $\partial$ is the border operator, through $\tilde{\delta}_{\mu\nu}(x;S_E)\mapsto\tilde{\delta}_{\mu\nu}
(x;S_E')=\tilde{\delta}_{\mu\nu}(x;S_E)+\partial_\mu\tilde{\delta}_\nu(x;V)-\partial_\nu\tilde{\delta}_\mu(x;V),\,\partial
V=S_E\cup S_E'$, the theory is modified. In the sequel we shall approach this point carefully by using an extension of the
JTA we have presented in \cite{csm, dafdc}. The procedure is as follows.

The dual of the Maxwell action is given by:
\begin{equation}
*\bar{S}=\int d^4x\left[-\frac{1}{4}(\tilde{F}_{\mu\nu}-\tilde{\Lambda}_{\mu\nu}^E)^2-\tilde{A}^\mu\tilde{j}_\mu\right],
\label{eq:17}
\end{equation}
where the couplings are inverted relatively to the ones present in (\ref{eq:14}): here the dual vector potential
$\tilde{A}_\mu$ couples minimally to the monopoles and non-minimally to the electric charges.

We suppose that for the electric charges there are only a few fixed (external) worldlines $L_E$ while for the monopoles we
suppose that there is a fluctuating ensemble of closed worldlines $L_M$ that can eventually proliferate. The magnetic
current is written in terms of the magnetic Dirac brane as $\tilde{j}^\sigma=\frac{1}{2}\epsilon^{\sigma\rho\mu\nu}
\partial_\rho\Lambda^M_{\mu\nu}=\tilde{g}\delta^\sigma(x;L_M),\,L_M=\partial S_M$. In order to allow the monopoles to
proliferate we must give dynamics to their magnetic Dirac branes since the proliferation of them is directly related to the
proliferation of the monopoles and their worldlines. Thus we supplement the dual action (\ref{eq:17}) with a kinetic term
for the magnetic Dirac branes of the form $-\frac{\epsilon_c}{2}\tilde{j}_\mu^2$, which preserves the local symmetries of
the system. This is an activation term for the magnetic loops. Hence, the partition function associated to the extended dual
action reads:
\begin{align}
&Z^c:=\int\mathcal{D}\tilde{A}_\mu\,\delta[\partial_\mu\tilde{A}^\mu]
e^{i\int d^4x\left[-\frac{1}{4}(\tilde{F}_{\mu\nu}-\tilde{\Lambda}_{\mu\nu}^E)^2\right]}Z^c[\tilde{A}_\mu],
\label{eq:18}
\end{align}
where the Lorentz gauge has been adopted for the dual gauge field $\tilde{A}_\mu$ and the partition function for the brane
sector $Z^c[\tilde{A}_\mu]$ is given by,
\begin{align}
&Z^c[\tilde{A}_\mu]:=\sum_{\left\{L_M\right\}}\delta[\partial_\mu\tilde{j}^\mu]
\exp\left\{i\int d^4x\left[-\frac{\epsilon_c}{2}\tilde{j}_\mu^2+\tilde{j}_\mu\tilde{A}^\mu\right]
\right\},
\label{eq:19}
\end{align}
where the functional $\delta$-distribution enforces the closeness of the monopole worldlines.

Next, use is made of the Generalized Poisson's Identity (GPI) (see Appendix A of \cite{dafdc} for a detailed discussion on
the subject) in $d=4$:
\begin{equation}
\sum_{\left\{L_M\right\}}\delta[\eta_\mu(x)- \delta_\mu(x;L_M)]=\sum_{\left\{\tilde{V}\right\}}e^{2\pi i\int d^4x\,
\tilde{\delta}_\mu(x;\tilde{V})\eta^\mu(x)},
\label{eq:20}
\end{equation}
where $L_M$ is a 1-brane and $\tilde{V}$ is the 3-brane of complementary dimension. The GPI works as an analogue of the
Fourier transform: when the lines $L_M$ in the left-hand side of (\ref{eq:20}) proliferate, the volumes $\tilde{V}$ in the
right hand side become diluted and vice versa. We shall say that the branes $L_M$ and $\tilde{V}$ (or the associated
currents $\delta_\mu(x;L_M)$ and $\tilde{\delta}_\mu(x;\tilde{V})$) are Poisson-dual to each other. Using (\ref{eq:20}) we
can rewrite (\ref{eq:19}) as:
\begin{align}
Z^c[\tilde{A}_\mu]&=\int\mathcal{D}\eta_\mu\,\sum_{\left\{L_M\right\}}
\delta\left[\tilde{g}\left(\frac{\eta_\mu}{\tilde{g}}-\delta_\mu(x;L_M)\right)\right]\nonumber\\
&\delta\left[\tilde{g}\left(\partial_\mu\frac{\eta_\mu}{\tilde{g}}\right)\right]\exp\left\{
i\int d^4x\left[-\frac{\epsilon_c}{2}\eta_\mu^2+\eta_\mu\tilde{A}^\mu\right]\right\}\nonumber\\
&=\int\mathcal{D}\eta_\mu\,\sum_{\left\{\tilde{V}\right\}}
e^{2\pi i\int d^4x\,\tilde{\delta}_\mu(x;\tilde{V})\frac{\eta^\mu}{\tilde{g}}}\int
\mathcal{D}\tilde{\theta}\nonumber\\
& e^{i\int d^4x\,\tilde{\theta}\partial_\mu\frac{\eta^\mu}{\tilde{g}}}
\exp\left\{i\int d^4x\left[-\frac{\epsilon_c}{2}\eta_\mu^2+\eta_\mu\tilde{A}^\mu\right]\right\}
\nonumber\\
&=\sum_{\left\{\tilde{V}\right\}}\int\mathcal{D}
\tilde{\theta}\,\int\mathcal{D}\eta_\mu\,\exp\left\{i\int d^4x\left[
-\frac{\epsilon_c}{2}\eta_\mu^2+\right.\right.\nonumber\\
&\left.\left.-\eta^\mu\frac{1}{\tilde{g}}
(\partial_\mu\tilde{\theta}-\tilde{\theta}_\mu^V-\tilde{g}\tilde{A}_\mu)\right]\right\},
\label{eq:21}
\end{align}
where we defined the Poisson-dual current $\tilde{\theta}_\mu^V:=2\pi\tilde{\delta}_\mu(x;\tilde{V})$.

Integrating the auxiliary field $\eta_\mu$ in the partial partition function (\ref{eq:21}) and substituting the result back
in the complete partition function (\ref{eq:18}) we obtain, as the effective action for the condensed phase in the dual
picture, the London limit of the $U(1)$ Dual Abelian Higgs Model (DAHM):
\begin{align}
*\bar{S}^L_{DAHM}&=\int d^4x\left[-\frac{1}{4}(\tilde{F}_{\mu\nu}-\tilde{\Lambda}_{\mu\nu}^E)^2+
\frac{m_{\tilde{A}}^2}{2\tilde{g}^2}(\partial_\mu\tilde{\theta}-\tilde{\theta}_\mu^V+\right.\nonumber\\
&\left.-\tilde{g}\tilde{A}_\mu)^2\right],
\label{eq:22}
\end{align}
where we defined $m_{\tilde{A}}^2:=\frac{1}{\epsilon_c}$. The Poisson-dual current, $\tilde{\theta}_\mu^V$, appears in
(\ref{eq:22}) as a vortex-like defect for the scalar field $\tilde{\theta}$ describing the magnetic condensate in the
dual picture, being a parameter that controls the monopole condensation \cite{dafdc}.

Next we are going to dualize this result and one could be concerned with the fact that (\ref{eq:22}) constitutes a
nonrenormalizable theory, thus requiring a cutoff in order to be well defined as an effective quantum theory. However,
one can always think of its UV completion, in this case the complete DAHM, which is renormalizable, and then take its
dual, taking the London limit afterwards \cite{ripka}. At least in the case considered here, the result is exactly the
same one obtains by directly dualizing the London limit (\ref{eq:22}) of the DAHM, thus justifying the procedure we shall
adopt in the sequel.

The dual action to (\ref{eq:22}) is given by \cite{dafdc}:
\begin{align}
\bar{S}_{eff}^V[\tilde{Y}_{\mu\nu}]&=\int d^4x\left[-\frac{1}{2}(\partial_\mu\tilde{Y}^{\mu\nu})^2
+\frac{m_Y^2}{4}\tilde{Y}_{\mu\nu}^2+\frac{m_Y}{2}\tilde{Y}_{\mu\nu}\tilde{L}_E^{\mu\nu}\right],
\label{eq:23}
\end{align}
where we have identified the phenomenological parameters $m_{\tilde{A}}\equiv m_Y$ and defined the electric brane invariant:
\begin{equation}
\tilde{L}_E^{\mu\nu}:=\frac{g}{2}(\tilde{\delta}^{\mu\nu}(x;S_E)+\partial^\mu\tilde{\delta}^\nu(x;\tilde{V})-
\partial^\nu\tilde{\delta}^\mu(x;\tilde{V})),
\label{eq:24}
\end{equation}
where we have also used the non-abelian version of the Dirac quantization condition to write $\frac{2\pi}{\tilde{g}}=
\frac{g}{2}$. $\tilde{L}_E^{\mu\nu}$ is an electric brane invariant provided we have $\tilde{\delta}^\mu(x;\tilde{V})
\mapsto\tilde{\delta}^\mu(x;\tilde{V}')=\tilde{\delta}^\mu(x;\tilde{V})-\tilde{\delta}^\mu(x;V),\,\partial V=S_E\cup
S_E'$ under the deformation $S_E\mapsto S_E',\,\partial S_E=\partial S_E'$ of the electric Dirac branes.

In fact, the complete form of the electric brane transformation, $S_E\mapsto S_E',\,\partial S_E=\partial S_E',\,\partial V=S_E\cup S_E'$, is given by (see equations (\ref{eq:17}) and (\ref{eq:22})):
\begin{equation}
\left\{\begin{array}{l}
\tilde{\delta}_{\mu\nu}(x;S_E)\mapsto\tilde{\delta}_{\mu\nu}(x;S_E')=\tilde{\delta}_{\mu\nu}(x;S_E)+\partial_\mu
\tilde{\delta}_\nu(x;V)+\\
\,\,\,\,\,\,\,\,\,\,\,\,\,\,\,\,\,\,\,\,\,\,\,\,\,\,\,\,\,\,\,\,\,\,\,\,\,\,\,\,\,\,\,\,\,\,\,\,\,\,\,\,\,\,\,
\,\,\,\,\,\,\,\,\,\, -\, \partial_\nu\tilde{\delta}_\mu(x;V),\\
\tilde{\delta}_\mu(x;\tilde{V})\mapsto\tilde{\delta}_\mu(x;\tilde{V}')=\tilde{\delta}_\mu(x;\tilde{V})-
\tilde{\delta}_\mu(x;V),\\
\tilde{A}_\mu\mapsto\tilde{A}_\mu '=\tilde{A}_\mu+\frac{g}{2}\tilde{\delta}_\mu(x;V).
\end{array}\right.
\label{eq:25}
\end{equation}

Equation (\ref{eq:23}) is the generalization of the equation (\ref{eq:16}) compatible with the local electric brane
symmetry corresponding to the freedom of deforming the unphysical electric Dirac strings through the spacetime without
modifying the physics. This result was first obtained by some of us in \cite{dafdc} and it is in consonance with the impossibility of spontaneously breaking local symmetries, a fact widely known as the Elitzur's theorem \cite{elitzur}. Since we have electric brane symmetry in the diluted phase \cite{csm, dafdc, mvf} it must be preserved also in the condensed phase.
Indeed, just like the local gauge symmetry implies the current conservation, the local brane symmetry implies the charge
quantization. If one of these local symmetries could really be broken, there would be no current conservation or charge
quantization in the broken phase, a fact that is not observed in Nature. The explanation is again the fact that a local
symmetry is really never broken \cite{elitzur}. However, notice that the brane symmetry is hidden in the electric brane
invariant $\tilde{L}_E^{\mu\nu}$ and it is this hidden realization of the brane symmetry that is called the ``spontaneous
breaking of the brane symmetry" \cite{csm, dafdc, mvf}.

To see wether this effective theory gives us chromoeletric confinement or not we must integrate the field of the monopole
condensate, $Y_{\mu\nu}$, in order to obtain the effective action describing the interaction between the electric currents
in the condensed phase. 

Integrating the Kalb-Ramond field in the partition function we obtain the following effective action describing
the interaction between prescribed electric currents immersed in the monopole condensate (see section 3.8.1 of
\cite{ripka}):
\begin{align}
\bar{S}^V_{eff}&=\int d^4x \,\left[ - \frac 1 4 (\tilde{L}^E_{\mu\nu})^2 - \frac 1 2 \partial_\mu
\tilde{L}_E^{\mu\nu} \frac{1}{\partial^2 + m_Y^2} \partial^\alpha\tilde{L}^E_{\alpha\nu}\right].
\label{eq:26}
\end{align}

Noticing that we can rewrite the electric current $j_\mu$ in terms of the electric brane invariant $\tilde{L}_{\mu\nu}^E$ as
$j_\mu=\frac{1}{2}\epsilon_{\mu\nu\alpha\beta}\partial^\nu\tilde{L}_E^{\alpha\beta}$, it can be shown that:
\begin{align}
\partial_\mu\tilde{L}_E^{\mu\nu}\frac{1}{\partial^2+m_Y^2}\partial^\alpha\tilde{L}^E_{\alpha\nu}&=
j_\mu\frac{1}{\partial^2+m_Y^2}j^\mu-\frac{1}{2}(\tilde{L}^E_{\mu\nu})^2+\nonumber\\
&+\frac{m_Y^2}{2}\tilde{L}^E_{\mu\nu}\frac{1}{\partial^2+m_Y^2}\tilde{L}_E^{\mu\nu}.
\label{eq:27}
\end{align}

Substituting (\ref{eq:27}) in (\ref{eq:26}) it is a simple algebraic task to show that:
\begin{equation}
\bar{S}^V_{eff}=\int d^4x\left[-\frac{m_Y^2}{4}\tilde{L}^E_{\mu\nu}\frac{1}{\partial^2+m_Y^2}\tilde{L}_E^{\mu\nu}
-\frac{1}{2}j_\mu\frac{1}{\partial^2+m_Y^2}j^\mu\right].
\label{eq:28}
\end{equation}

The first term in (\ref{eq:28}) is responsible for the charge confinement: it ``spontaneously breaks the electric brane
symmetry" such that the electric brane invariant $\tilde{L}_{\mu\nu}^E$ acquires energy and constitutes the electric flux
tube connecting two charges of opposite sign immersed in the monopole condensate. The flux tube has a thickness equal to
the penetration depth of the electric field in the dual superconductor constituted by the magnetic condensate. The shape of
the electric flux tube that corresponds to the stable configuration that minimizes the energy of the system is that of a
straight tube (minimal space). Substituting in the first term of (\ref{eq:28}) such a solution for the brane invariant,
$\tilde{L}_{\mu\nu}^E=\frac{1}{2}\epsilon_{\mu\nu\alpha\beta}\frac{1}{n\cdot\partial}(n^\alpha j^\beta-n^\beta j^\alpha)$,
where $n^\mu:=(0,\v R:=\v R_1-\v R_2)$ is a straight line connecting the electric charges $+\frac{g}{2}$ in $\v R_1$ and
$-\frac{g}{2}$ in $\v R_2$, and taking the static limit we obtain a linear confining potential between the electric charges
\cite{ripka}.

The second term in (\ref{eq:28}) describes the Yukawa-like short-range interaction between the electric currents in the
condensed phase.

Taking the limit $m_Y\rightarrow 0$ leads us back to the diluted phase, eliminating the monopole condensate and destroying
the confinement. Indeed, we can see that in this limit the interaction between the electric currents in (\ref{eq:28})
becomes of the long-range (Coulomb) type and the confining term goes to zero.

It is very important to make a final remark regarding the JTA. In \cite{dafdc} we made the observation that under a complete
monopole condensation (\textit{id est}, when we consider that the monopoles proliferate until occupy the whole space) we
have $\tilde{\theta}_\mu^V\rightarrow 0$ and we recover from (\ref{eq:23}) the Quevedo-Trugenberger result (\ref{eq:16}).
However, as we discussed here, (\ref{eq:16}) is incompatible with the local electric brane symmetry and since a local
symmetry can not be broken \cite{elitzur}, (\ref{eq:16}) must be substituted by (\ref{eq:23}). The fact is that it is
impossible to have a complete monopole condensation when we include external electric charges in the system since the
electric fields generated by them, although expulsed of almost all the space by the dual Meissner effect, can not simply
vanish: they are confined into straight flux tubes connecting electric charges of opposite sign immersed in the monopole
condensate. These vortices with opposite electric charges in their borders do not vanish (only vortices disconnected from
the electric charges can vanish) and thus there is no complete monopole condensation when there are external electric
charges immersed in the dual superconductor. Furthermore, the physics described by (\ref{eq:23}) features not only the
electric charge confinement but also the charge quantization since the brane symmetry is maintained in the condensed phase.
The scenario is quite different when we consider (\ref{eq:16}), where although the electric charge confinement is present,
the charge quantization is lost due to the explicitly breaking of the brane symmetry. To have the right physics with
electric charge confinement and charge quantization in the condensed phase we must be very carefull and give a proper
treatment of brane symmetry as we did in this section.

Kleinert was the first one to point out that the brane symmetry is a kind of local symmetry different from the gauge
symmetry \cite{mvf}. We generalized the JTA \cite{JT} as done in \cite{csm, dafdc}, making it compatible with the Elitzur's theorem and the local brane symmetry.

\section{Conclusion}
\label{sec:conclusion}

In this Letter we used the Julia-Toulouse condensation mechanism \cite{JT}, as generalized by some of us in \cite{csm, dafdc},
to study the confinement problem for a $SU(2)$ gauge theory.

We took as the starting point to the novel reasoning presented in this Letter to approach the monopole condensation, 
in the non-abelian case, the expression for the restricted $SU(2)$ gauge theory defined by means of the Cho decomposition
of the non-abelian connection. We showed that, in the magnetic gauge, the action can be put in the form of the Maxwell
theory minimally coupled to external chromoelectric charges and non-minimally coupled to chromomagnetic monopoles. This
was the crucial point that allowed us to apply the Generalized JTA for defects condensation and obtain an effective theory
compatible with Elitzur's Theorem and local electric brane symmetry for the phase where the monopoles are condensed.

In order to obtain the physics describing the interaction between external chromoeletric charges immersed in the magnetic
condensate we integrated out in the partition function the field of the monopole condensate. The effective action found
displays an Yukawa short-range interaction between the electric currents in the condensed phase and a term responsible for
the confinement physics at large distances, giving a linear potential in the interquarks separation when we consider the
static case. Furthermore, since our generalized approach to the JTA preserves the local electric brane symmetry, the charge
quantization that is present in the diluted phase is maintained in the condensed phase.

The result here achieved also confirms that the restricted gauge theory proposed by Cho is indeed the subsector of the
complete gauge theory responsible for the confinement physics.

It is also important to say that the decomposition (\ref{eq:2}) was also approached in a different way by Faddeev and
Niemi \cite{faddeev}. The difference between Cho's approach and Faddeev-Niemi's approach regards the specific form of the
valence potential, $\v{X}_\mu$. In Cho's construction the field $\hat{n}$ is regarded as a topological variable and its
2 degrees of freedom are not counted as transverse modes for the gluons - in doing so, the valence potencial in Cho's
interpretation of the decomposition (\ref{eq:2}) carries 4 transverse modes, being the other 2 transverses modes carried
by the abelian component, $A_\mu$ \cite{cho}. On the other hand, Faddeev and Niemi interpret the 2 degrees of freedom of
$\hat{n}$ as being 2 of the 6 transverse modes of the gluons and in doing so, their valence potential has only 2 transverse
modes, the other 2 transverse modes being carried by $A_\mu$ \cite{faddeev}. The total number of degrees of freedom present
in the connection described by (\ref{eq:2}) after gauge fixing in Cho's approach is 8 (6 transverse modes + 2 topological
modes) while in Faddeev-Niemi's approach is 6 (6 transverse modes). Thus, in Faddeev-Niemi's approach the number of physical
degrees of freedom of the $SU(2)$ connection (in 3+1 there are 6 of them) is preserved by the decomposition, while in Cho's
approach it is not (in the last of the references in \cite{cho}, Cho discusses that his interpretation of the decomposition
(\ref{eq:2}) indeed modifies the quantum theory). Regarding our result, as we did not specify the form of our valence
potential, since we discharged it in our discussion, we expect that it remains unchanged in either approach (Cho or
Faddeev-Niemi), since both of them agree about the form of the restricted connection that was the essential element we used
in this Letter.

More recently, Faddeev and Niemi proposed a novel decomposition of the $SU(2)$ connection in terms of spin-charge separated
variables \cite{fn} constructed directly in terms of the components of the non-abelian connection. The lowest order
effective Lagrangian density expressed in terms of the spin-charge separated variables is given by eq. (60) of \cite{fn}.
To make contact with our results we notice that we can recover the functional form of the restricted gauge theory from the
complete gauge theory by setting $\rho=0$ in eq. (60) of \cite{fn}.

It is further claimed in \cite{fn} that the condition $\rho=\textrm{constant}$ is related with the non-perturbative
contribution of the $\langle A^2\rangle$ condensate \cite{zakharov}. It would be interesting to have a better understanding
of the interplay between this condensate and the monopole condensate studied in the present Letter.

\section{Acknowledgements}

We thank Conselho Nacional de Desenvolvimento Cient\'ifico e Tecnol\'ogico (CNPq) for financial support.

\end{document}